**On the resilience of magic number theory for conductance ratios of aromatic molecules.**

Lara Ulčakar[1], Tomaž Rejec[2,1], Jure Kokalj[3,1], Sara Sangtarash[*], Hatef Sadeghi[*], Anton Ramšak[2,1], John H. Jefferson[*] and Colin J. Lambert[*]

[*] Department of Physics, Lancaster University, Lancaster, LA1 4YB, United Kingdom.

[1] Jožef Stefan Institute, Ljubljana, Slovenia.

[2] Faculty of Mathematics and Physics, University of Ljubljana, Ljubljana, Slovenia.

[3] Faculty of Civil and Geodetic Engineering, University of Ljubljana, Ljubljana, Slovenia.

Corresponding author: c.lambert@lancaster.ac.uk

**ABSTRACT**

**If simple guidelines could be established for understanding how quantum interference (QI) can be exploited to control the flow of electricity through single molecules, then new functional molecules, which exploit room-temperature QI could be rapidly identified and subsequently screened. Recently it was demonstrated that conductance ratios of molecules with aromatic cores, with different connectivities to electrodes, can be predicted using a simple and easy-to-use "magic number theory." In contrast with counting rules and "curly-arrow" descriptions of *destructive* QI, magic number theory captures the many forms of *constructive* QI, which can occur in molecular cores. Here we address the question of how conductance ratios are affected by electron-electron interactions. We find that due to cancellations of opposing trends, when Coulomb interactions and screening due to electrodes are switched on, conductance ratios are rather resilient. Consequently, qualitative trends in conductance ratios of molecules with extended pi systems can be predicted using simple "non-interacting" magic number tables, without the need for large-scale computations. On the other hand, for certain connectivities, deviations from non-interacting conductance ratios can be significant and therefore such connectivities are of interest for probing the interplay between Coulomb interactions, connectivity and QI in single-molecule electron transport.**

Introduction

Understanding and exploiting room-temperature quantum interference (QI) in single molecules is the key to creating new high-performance single-molecule devices and thin-film materials formed from self-assembled molecular layers. During the past decade, experimental and theoretical studies of single molecules attached to metallic electrodes have demonstrated that room-temperature electron transport is controlled by QI within the core of the molecule [1-20]. Many of these demonstrations have been achieved by noting that in contrast with artificial quantum dots, where atomic-scale details of the coupling of a dot to external electrodes are not known, the connectivity to the core of a single molecule may be controlled to atomic accuracy. Fig. 1 shows two examples of molecules with a common anthanthrene core, connected via triple bonds and pyridyl anchor groups to gold electrodes. The anthanthrene core (represented by a lattice of 6 hexagons) of molecule **1** and the anthanthrene core of molecule **2** are connected differently to the triple bonds. Therefore it is natural to ask how the electrical conductance and interference properties of such molecules are affected by connectivity.

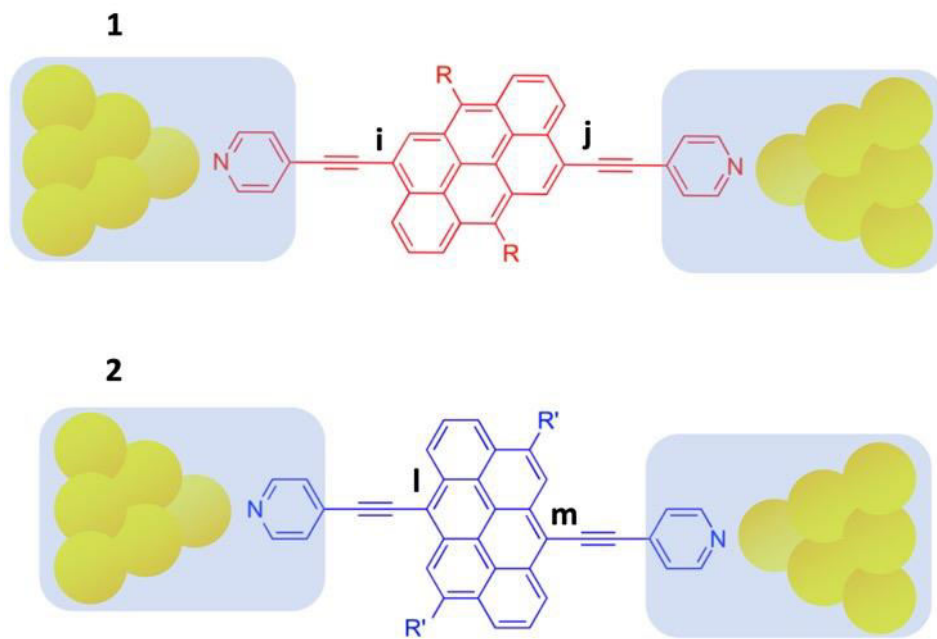

**Figure 1.** Examples of molecules with anthanthrene cores, connected via triple bonds and pyridyl anchor groups to the tips of gold electrodes, which in turn connect to crystalline gold leads (not shown). Molecule **1** has a connectivity *i-j* and electrical conductance $\sigma_{ij}$, while molecule **2** has a connectivity *l-m* and electrical conductance $\sigma_{lm}$.

………………………………………………………………………………………………………………

In a typical experiment using mechanically controlled break junctions or STM break junctions [13-18], fluctuations and uncertainties in the coupling to electrodes are dealt with by measuring the conductance of such molecules many thousands of times and reporting the statistically-most-probable electrical conductance, just before the junction breaks. If $\sigma_{ij}$ is the statistically-most-probable conductance of a molecule such as **1**, with connectivity *i-j* and $\sigma_{lm}$ is the corresponding conductance of a molecule such as **2**, with connectivity *l-m*, then it was recently predicted theoretically and demonstrated experimentally [21-23] that for polyaromatic hydrocarbons (PAHs) such as anthanthrene, the statistically-most-probable conductance ratio $\sigma_{ij}/\sigma_{lm}$ is independent of the coupling to the electrodes and could be obtained from tables of "magic numbers," which for bipartite PAHs in the absence of electron-electron interactions, are simply tables of integers. If $M_{ij}$ ($M_{lm}$) is the magic number corresponding to connectivity *i-j* (*l-m*), then this "magic ratio theory" predicts

$$\frac{\sigma_{ij}}{\sigma_{lm}} = \left(\frac{M_{ij}}{M_{lm}}\right)^2. \tag{1}$$

From a conceptual viewpoint, magic ratio theory views the shaded regions in Fig. 1 as "compound electrodes", comprising both the anchor groups and gold electrodes, and focuses attention on the contribution from the core alone. The validity of Eq. (1) rests on the following key foundational concepts [1,2, 21-23]:
1. weak coupling
2. locality
3. connectivity
4. mid-gap transport
5. phase coherence
6. connectivity-independent statistics

When these conditions apply, the complex and often uncontrolled contributions from electrodes and electrode-molecule coupling cancel in conductance ratios and therefore a theory of conductance ratios can be developed by focussing on the contribution from molecular cores alone.

The term "weak coupling" means that the central aromatic subunit such as anthanthrene should be weakly coupled to the anchor groups *via* spacers such as acetylene. "Locality" means that when a current flows through an aromatic subunit, the points of entry and exit are localised in space. For example in molecule **1**, the current enters at a particular atom *i* and exits at a particular atom *j*. The concept of "connectivity" recognises that through chemical design, spacers can be attached to different parts of a central subunit with atomic accuracy and therefore it is of interest to examine how the flow of electricity depends on the choice of connectivity to the central subunit. The concept of "mid-gap transport" is recognition of the fact that unless a molecular junction is externally gated by an electrochemical environment or an electrostatic gate, charge transfer between the electrodes and molecule ensures that the energy levels adjust such that the Fermi energy $E_\text{F}$ of the electrodes is usually located in the vicinity of the centre of the HOMO-LUMO gap and therefore transport takes place in the co-tunnelling regime. In other words, transport is usually "off-resonance". The concept of "phase coherence" recognises that in this co-tunnelling regime, the phase of electrons is usually preserved as they pass through a molecule and therefore transport is controlled by QI. The condition of "connectivity-independent statistics" means that the statistics of the coupling between the anchor groups and electrodes should be independent of the connectivity to the aromatic core. When each of these conditions applies, it can be shown [1,2,21,22] that the most probable electrical conductance corresponding to connectivity *i,j* is proportional to $\left|G_{ij}(E_\text{F})\right|^2$ where $G_{ij}(E_\text{F})$ is the Green's function of the core alone, evaluated at the Fermi energy of the electrodes. In the absence of time-reversal symmetry breaking, $G_{ij}(E_\text{F})$ is a real number. Since only conductance ratios are of interest, we define magic numbers by

$$M_{ij} = AG_{ij}(E_\text{F}), \tag{2}$$

where *A* is an arbitrary constant of proportionality, chosen to simplify magic number tables and which cancels in Eq. (1). Magic ratio theory applies to any single-molecule junction, provided conditions 1-6 are satisfied. It represents an important step forward, because apart from the Fermi energy $E_\text{F}$, no information about the electrodes is required. The question we address below is what are the precise values of the numbers $M_{ij}$ and how are they affected by electron-electron interactions?

In the literature, several papers discuss the conditions for *destructive* QI, for which $M_{ij} \approx 0$ [6,9-18, 24-29]. On the other hand, magic ratio theory aims to describe *constructive* QI, for which $M_{ij}$ may take a variety of non-zero values. If *H* is the non-interacting Hamiltonian of the core, then since the matrix $G(E_\text{F}) = (E_\text{F} - H)^{-1}$, the magic number table is obtained from a matrix inversion, whose size and complexity reflects the level of detail contained in *H*. The quantities $M_{ij}$ were termed "magic" [21-23], because even a simple theory based on connectivity alone yielded values, which were found to be in remarkable agreement with experiment. For example for molecule **1**, the prediction was $M_{ij} = -1$, whereas for molecule **2**, $M_{lm} = -9$ and therefore the electrical conductance of **2** was predicted to be 81 times higher than that of **1**, which is close to the measured value of 79. This large ratio is a clear manifestation of quantum interference (QI), since such a change in connectivity to a classical resistive network would yield only a small change in conductance. To obtain the above values for $M_{ij}$ and $M_{lm}$, the Hamiltonian *H* was chosen to be

$$H = \begin{pmatrix} 0 & C \\ C^t & 0 \end{pmatrix}, \qquad (3)$$

where the connectivity matrix $C$ of anthanthrene is shown in Fig. 2b. In other words, each element $H_{ij}$ was chosen to be -1 if $i,j$ are nearest neighbours or zero otherwise and since anthanthrene is represented by the bipartite lattice in which odd numbered sites are connected to even numbered sites only, $H$ is block off diagonal. The corresponding core Green's function evaluated at the gap centre $E_\text{F} = 0$ is therefore obtained from a simple matrix inversion $G(0) = -H^{-1}$. Since $H$ and therefore $-H^{-1}$ are block off-diagonal, this yields the following structure for the magic number table of the PAH core $M = \begin{pmatrix} 0 & \bar{M}^t \\ \bar{M} & 0 \end{pmatrix}$. The off-diagonal block of the magic number table $\bar{M}$ for anthanthrene is shown in Fig. 2c. As noted above, for molecule **1**, with connectivity 9-22, $M_{9,22} = -1$, whereas for molecule **2**, with connectivity 3-12, $M_{3,12} = -9$.

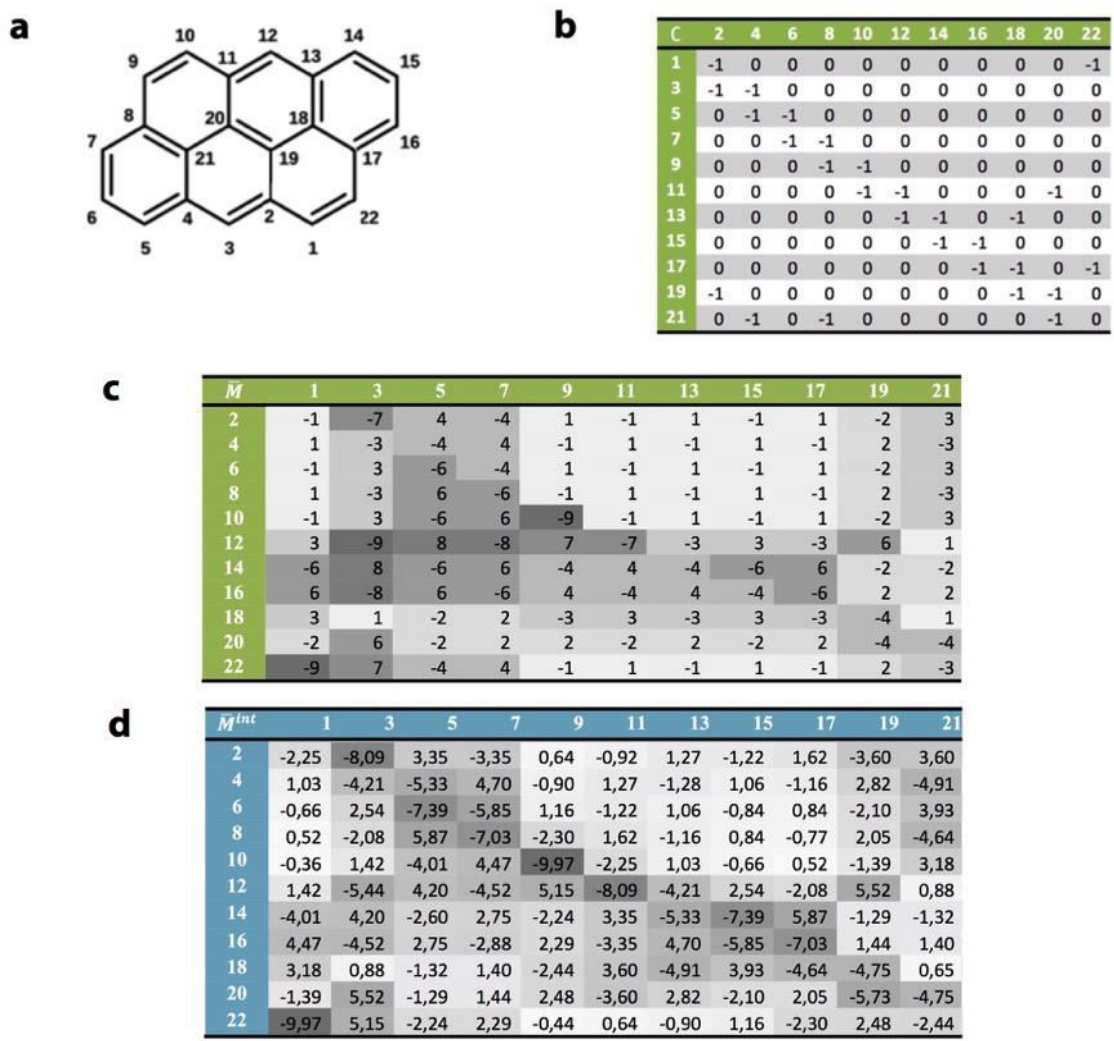

**Figure 2.** (a) The numbering system for pi orbitals of an anthanthrene core. (b) The connectivity table $C$ for anthanthrene. (c) The non-interacting magic number table $\bar{M}$ corresponding to the anthanthrene lattice (a). (d) The interacting magic number table $\bar{M}^\text{int}$ corresponding to the anthanthrene lattice in the presence of electron-electron interactions, calculated within the Hartree-Fock approximation. The depth of shading in the tables is in proportion to the table entries and highlights the qualitative agreement between the non-interacting and interacting magic number tables.

………………………………………………………………………………………………………………………

Magic number tables such as Fig. 2c are extremely useful, since they facilitate the identification of molecules with desirable conductances for future synthesis. Conceptually, tables obtained from Hamiltonians such as Eq. (3) are also of interest, since they capture the contribution from intra-core connectivity alone (via the matrix $C$, comprising -1's or zeros), while avoiding the complexities of chemistry. Although magic number tables obtained from such connectivity matrices were shown to agree qualitatively with break junction measurements of several different molecules carried out by different experimental groups [22], the errors in the experimental estimates of conductance ratios are rather large and the number of molecules tested is small. Therefore it is of interest to seek to improve the accuracy of magic number tables by utilising more accurate core Hamiltonians. An essential ingredient missing from the Hamiltonian of Eq. (3) is electron-electron interactions and therefore in what follows we aim to obtain improved estimates of magic numbers by including the effect of Coulomb interactions and screening. Results will be presented for a variety of graphene-like molecules, including benzene, naphthalene, anthracene, pyrene and anthanthrene.

The main outcome of this study is exemplified by the interacting magic number table $M^{\text{int}}$ for anthanthrene, whose lower off-diagonal block $\bar{M}^{\text{int}}$ is presented in Fig. 2d. Note that magic numbers are only defined up to a constant of proportionality, which does not affect the predicted conductance ratios. Therefore to facilitate comparison between interacting and non-interacting values, in the table of Fig. 2d, the constant is chosen to minimise the mean square deviation between the non-interacting and interacting M-tables. The latter shows for example, that in the presence of Coulomb interactions, the magic number for molecule **1**, changes from -1 to $M^{\text{int}}_{9,22} = -0.44$, whereas for molecule **2**, the magic number changes from -9 to $M^{\text{int}}_{3,12} = -5.44$. Hence interacting magic number theory predicts that the conductance of **2** is (-5.44/-0.44)²=152 times higher than that of **1** (or more precisely 148 if magic numbers to 3 decimal places are used, as presented in the Supplementary Information (SI)). This demonstrates that the conductance ratio of 81, predicted by non-interacting magic numbers is qualitatively correct (ie to within a factor of 2). Furthermore comparison between tables c and d in Fig. 2 shows that the non-interacting magic number table captures the qualitative trends of the interacting magic number table. This qualitative agreement is remarkable, since the former can be obtained from a few lines of e.g. MATLAB code, while the latter is the result of a substantial many-body calculation. Results for both non-interacting and interacting magic number tables of a range of PAHs are presented in the SI. Our main conclusion is that non-interacting magic numbers are a useful qualitative guide for predicting conductance ratios, even in the presence of Coulomb interactions and screening.

**Results**

In the following numerical simulations, the transmission coefficient $T_{ij}(E)$ describing the probability that electrons of energy $E$ can pass from one electrode to another via sites $i,j$. Systems with the chiral symmetry have a symmetric energy spectrum which means that for half-filled systems the Fermi energy is at the gap centre. Therefore, conductance ratios are obtained from $T_{ij}(0)$. To include the effects of the Coulomb interaction, we first generalise the Hamiltonian of Eq. (3) to the interacting Parr-Pariser-Pople (PPP) model [30-32]. We base our treatment of the Coulomb interaction on a scheme proposed by Ohno [33], which obtains inter-site interaction integrals by smoothly interpolating between the Hubbard integral $U$ for zero separation between sites and an unscreened Coulomb interaction for large separations between sites. This is an established model for the aromatic molecules and yet its simplicity enables us to study the effect of interaction. Recently it was shown experimentally [34] that molecular levels shift as a result of Coulomb interaction with image charges in the metal leads, resulting in a HOMO-LUMO gap renormalization. Therefore, we also take into account additional electric potential screening, which is induced by the conducting electrodes. We model the latter as infinite parallel plates located at a distance $d$ from each of the connection sites.

Calculations for smaller molecules (benzene, naphthalene, and anthracene) are performed using both the Lanczos exact diagonalization method [35] and using the restricted Hartree-Fock (HF) approximation (for technical details see Methods and the SI). We use the latter since we consider effects of the Coulomb interaction in the simplest scheme possible (for superior approximate methods as for example GW method see [36]). For the larger molecules (pyrene and anthanthrene) the Lanczos "calculation is not feasible. For the smaller molecules, where it is possible to compare the Lanczos method with the HF approximation, agreement was found for the HOMO-LUMO gaps (within approx. 1%) and conductances (within approx. 10%) for different connectivities. This gives us confidence that use of the HF approximation for the larger molecules is valid.

As a first example, we present results for the conductance ratio of molecules with naphthalene cores (see Fig. 3a), with two different connectivities, denoted 6-9 and 3-8, whose non-interacting magic ratio is 4 [see table in Fig. 3c of the SI]. To elucidate the effects of varying the strength of interactions, we multiply all the interaction integrals by a scale factor $\lambda$ and examine the effect of varying $\lambda$. The upper table in Fig. 3b shows a comparison between results obtained using HF and direct Lanczos diagonalization for different values of the scaling parameter $\lambda$, ranging from $\lambda = 0$ (non-interacting) to $\lambda = 1$ (interacting) and to the greater, unphysical value of $\lambda = 2$. For $\lambda = 1$, the lower table in Fig. 3b shows the effect of screening by electrodes at different distances $d$ from molecule, ranging from $d = d_0$, where $d_0$ is the carbon-carbon bond length, to $d = \infty$ (no screening). Fig. 3c shows that the HF approximation reproduces the exact Lanczos HOMO-LUMO gap correctly for naphthalene and while there is a small discrepancy in the transmission coefficient (Fig. 3d) at the Fermi level $E = 0$, the HF conductance ratio is qualitatively correct, deviating appreciably from the exact value only when $\lambda$ becomes much larger than the physically-relevant one. Note that the conductance ratio at $\lambda = 0$ is not exactly equal to the non-interacting ratio of 4 due to the presence of a small but finite coupling of the molecule to the electrodes.

The lower table in Fig. 3b shows that screening by the electrodes does not change the ratio appreciably even though the renormalization of the HOMO-LUMO gap is different for different connectivities. The difference in gap renormalisation occurs, because screening is more effective when the distance between electrodes is small. The gap is thus reduced more by screening for the 6-9 connectivity, where the long axis of the molecule is parallel to electrode surfaces, than for the 3-8 connectivity, where it is perpendicular to them. If the QI between different paths through the molecule did not change, one would expect the conductance ratio to be proportional to the ratio of inverse gaps squared and therefore the conductance ratio at $d/d_0 = 1$ should have increased by 31% compared to the conductance ratio in absence of screening. Here this effect is almost exactly compensated by the screening induced change of QI between different paths through the molecule.

The results in Fig. 3b show that for naphthalene, the HF and Lanczos predictions for the conductance ratio are rather close to each other and to that of (non-interacting) magic number theory. As a second example, Fig. 4b shows HF results for the conductance ratio of molecules **1** and **2** with anthanthrene cores and Fig. 4c shows their corresponding transmission functions $T(E)$. As for naphthalene, the conductance ratio increases from the non-interacting value when interactions are present (upper table in Fig. 4b), but here the deviations from the non-interacting magic ratio of 81 are more pronounced. In contrast with naphthalene, the conductance ratio is also affected by screening: when the electrodes become closer to the molecule, the ratio drops back towards the non-interacting value. In contrast with naphthalene, the rescaling of the HOMO-LUMO gaps of both connectivities would lead to an increase of the conductance ratio (by 37% for $d/d_0 = 1$), so the drop of the conductance ratio can be attributed to screening-induced change in the QI of different paths through the molecule.

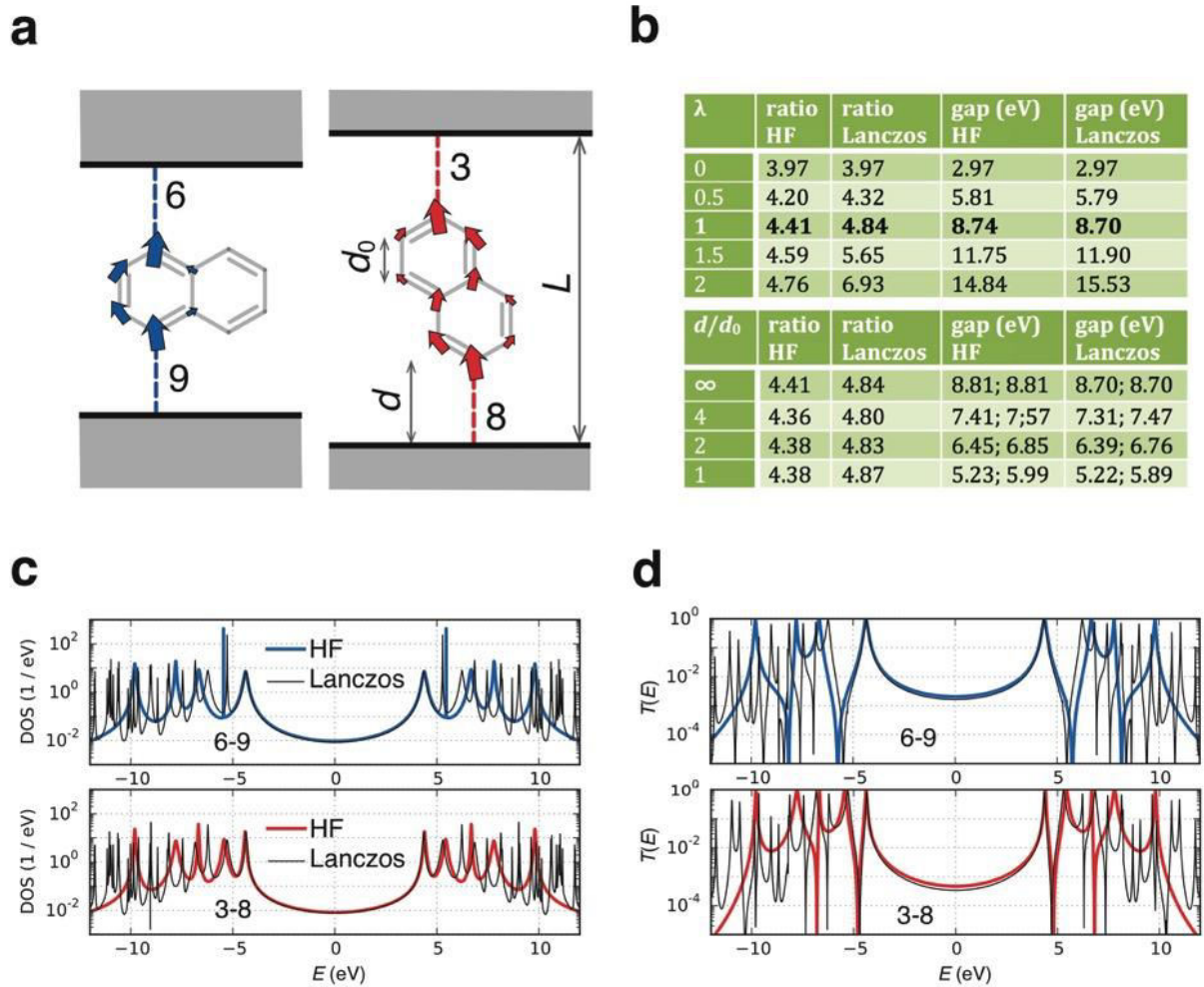

**Figure 3.** HF and Lanczos results for the naphthalene molecule within the PPP model. **a)** Naphthalene molecules with 6-9 (left) and 3-8 (right) connectivities to electrodes, whose non-interacting conductance ratio is predicted to be 4. Arrows show how currents are distributed in a molecule when a small source-drain bias is applied between the electrodes. **b, top)** HF and Lanczos conductance ratios (columns 2 and 3, respectively) and HOMO-LUMO gaps (columns 4 and 5, respectively) for different interaction strengths $\lambda$, with no screening by electrodes. **b, bottom)** HF and Lanczos conductance ratios for the physical value of $\lambda = 1$, in the presence of screening by electrodes at different distances $d$ away from the molecule, measured in units of the carbon-carbon bond length $d_0$. An infinite distance corresponds to no screening. The two values of the HOMO-LUMO gap separated by a semicolon correspond to the 6-9 and 3-8 connectivities, respectively. **c)** The density of states in the molecule for the 6-9 connectivity (top) and for the 3-8 connectivity (bottom). The coloured and the black line show the HF and the Lanczos result, respectively. **d)** The transmission function for the 6-9 connectivity (top) and for the 3-8 connectivity (bottom). The coloured and black lines show HF and the Lanczos result, respectively.
……………………………………………………………………………………………………………

To highlight the correlation (and differences) between non-interacting and interacting conductance ratios, the blue dots in Fig. 4d are plots of HF conductance ratios versus those predicted by non-interacting magic numbers for all possible pairs of connectivities and shows that there is a significant degree of correlation between the two. The main conclusion from these results and for corresponding results for other molecules (see SI) is that although Coulomb interactions and screening cause the conductance ratios to vary, in many cases the non-interacting magic ratios provide the correct qualitative trend. In the case of anthanthrene (Fig. 4), the non-interacting ratio of 81 is surprisingly close to the most-physical conductance ratio of 79.3, which occurs at $\lambda = 1$ and a screening distance of $d/d_0 = 1$.

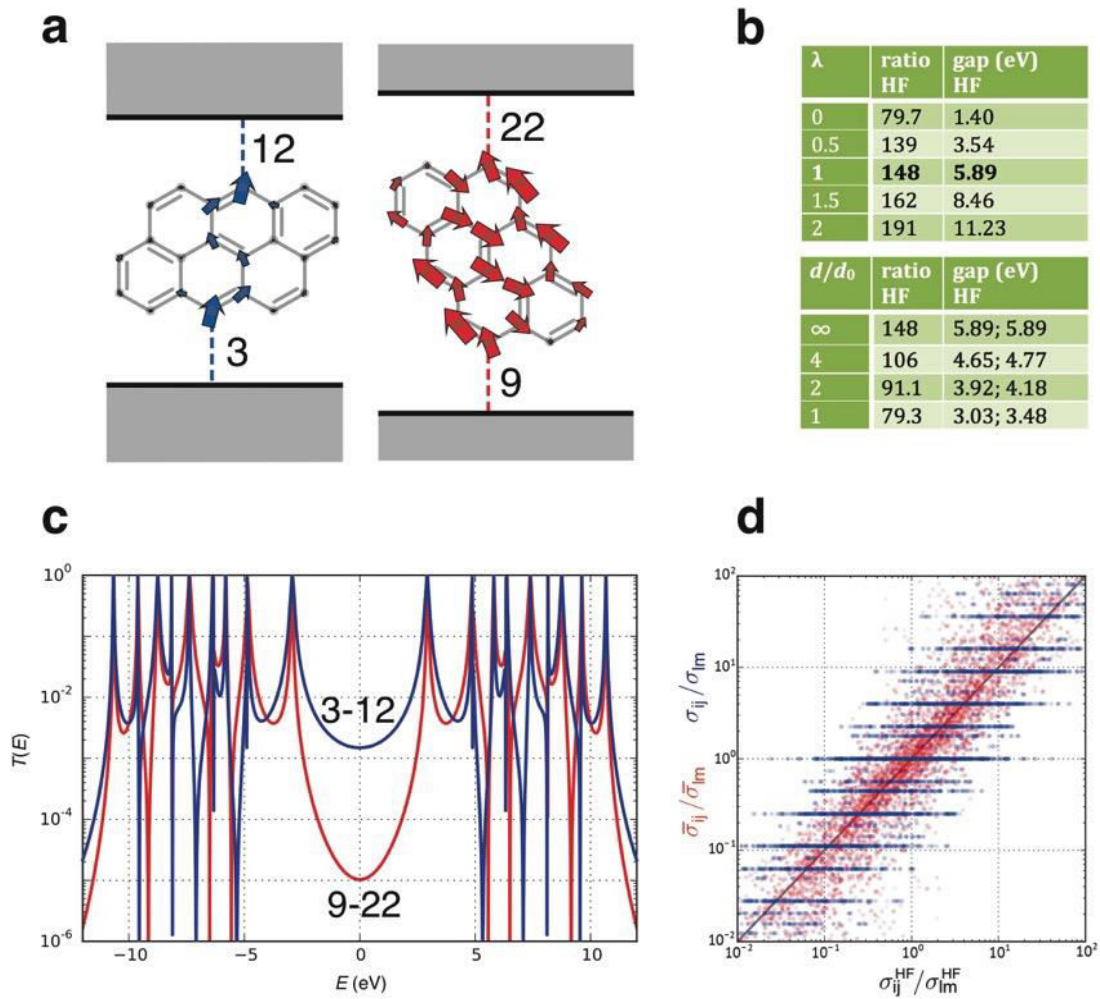

**Figure 4.** Results for the anthanthrene molecule within the PPP model. **a)** The anthanthrene molecule attached to electrodes for molecule **2** with 3-12 connectivity (left) and molecule **1** with 9-22 connectivity (right). Arrows as in Fig. 3a. **b)** As in Fig. 3b, but only HF results are tabulated here. **c)** Transmission functions for the 3-12 connectivity (blue) and for the 9-22 connectivity (red). **d)** Correlations of the HF conductance ratio (horizontal axis) for a particular pair of connectivities with the non-interacting (blue dots) and the infinite-range interaction (orange dots) conductance ratio for the same pair of connectivities. Results for all possible pairs of connectivities are shown.

………………………………………………………………………………………………………………

The above results are obtained from the PPP model, which coincides with the non-interacting Hamiltonian of Eq. (3) when $U = 0$. This model preserves chiral symmetry and guarantees that the centre of the HOMO-LUMO gap lies in the middle of the energy spectrum ($E = 0$). The model captures the effect of connectivity and Coulomb interactions, without introducing complexities associated with the chemical nature of the molecules. To include the latter, we used density functional theory to compute the transmission coefficient $T(E)$ of molecules with different connectivities attached to gold electrodes. Fig. 5a shows plots of $\log T(E)$ versus $E$ for the 3-8 and 6-9 connectivities of naphthalene and Fig. 5b shows $\log T(E)$ versus $E$ for the 3-12 and 9-22 connectivities of anthanthrene. To highlight the further role of chemistry, the bottom right inset of Fig. 5b shows corresponding results when the anthanthrene core is directly coupled to gold electrodes, as shown in the bottom left inset. For energies in the shaded regions of these plots, the ratio of geometric averages of transmission coefficients approximately coincides with the non-interacting magic ratio rule (see Table 1, column 7).

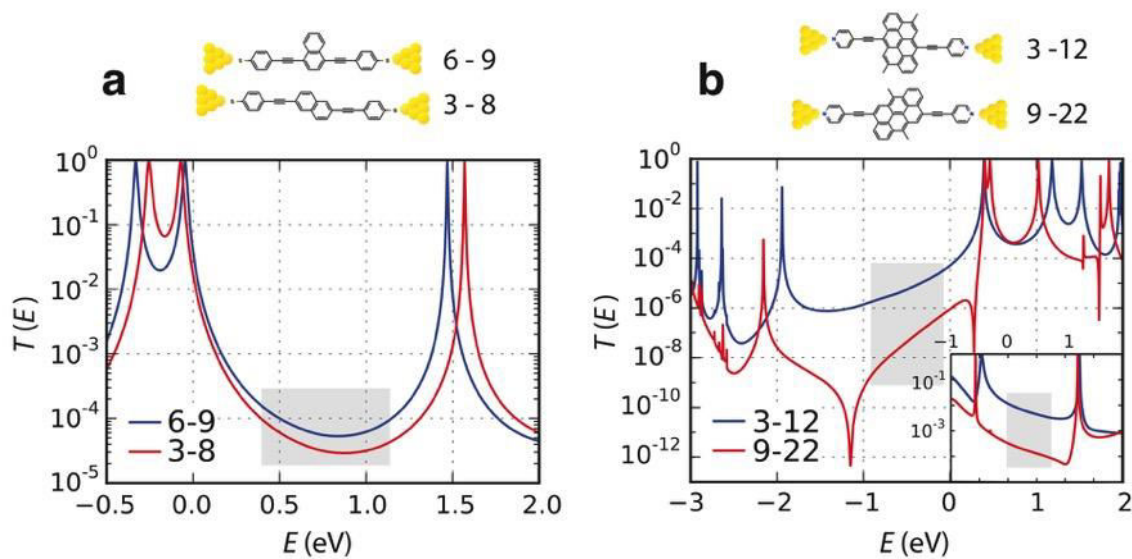

**Figure 5.** (a) DFT results for the transmission coefficients of naphthalene with 6-9 and 3-8 connectivities attached to gold electrodes; (b) DFT results for transmission coefficients of anthanthrene with 3-12 and 9-22 connectivities (molecules **1** and **2**) attached to gold electrodes. The bottom-right inset shows corresponding transmission coefficients when the anthanthrene cores are directly coupled to gold electrodes.

………………………………………………………………………………………………………………

PPP and DFT results for conductance ratios of benzene, anthracene and pyrene with two connectivities are presented in Figs. 7-12 of the SI. Except for benzene, the conductance ratios for those connectivities were measured experimentally. Table 1 shows a comparison between these results, the non-interacting magic ratios and experiment.

| Molecule ($i$-$j$; $l$-$m$) | Non-interacting ratio | PPP (HF) | PPP (Lanczos) | Infinite range interaction | PPP $d/d_0 = 1$ (HF) | DFT | Experiment |
|---|---|---|---|---|---|---|---|
| benzene (1-2; 1-4) | 1 | 1.39 | 1.60 | 1.75 | - | - | - |
| naphthalene (6-9; 3-8) | 4 | 4.41 | 4.84 | 4.59 | 4.38 | 2 | 5.1 [15] |
| anthracene (1-8; 5-12) | 16 | 21.2 | 26.3 | 19.8 | 22.0 | 13 | 10.2 [15] |
| pyrene (5-12; 4-11) | 9 | 3.92 | - | 4.65 | 5.78 | 9 | 8 [21] |
| anthanthrene (3-12; 9-22) | 81 | 148 | - | 2400 | 79.3 | 81 | 79 [22] |

**Table 1.** Conductance ratios for various molecules, for a pair of connectivities listed in the first column. Column 2: Ratios obtained using non-interacting magic number theory. Column 3: PPP with no screening by electrodes, calculated within the HF approximation. Column 4: PPP with no screening by electrodes, calculated using the Lanczos method, where available. Column 5: PPP with an infinite range interaction. Column 6: PPP with screening by electrodes at $d/d_0 = 1$, within the HF approximation. Column 7: DFT geometric average ratio over the shaded regions of figure 5. Column 8: experimental ratios, where available.

………………………………………………………………………………………………

The effect of interaction on conductance ratios can be roughly estimated by considering the PPP model in the infinite-range interaction limit (where the interaction integrals take the same value $\tilde{U}$ for all pairs of sites in a molecule), which can be solved exactly for an isolated molecule. In this limit, the core Green's function takes the form $\tilde{G}(0) = -\left(H + \frac{1}{2}\tilde{U}\text{sgn}H\right)^{-1}$ and can be easily evaluated from the connectivity matrix $C$ and $\tilde{U}$ alone, with $H = \begin{pmatrix} 0 & C \\ C^t & 0 \end{pmatrix}$ (see Supplementary Information). For $\tilde{U}$ we take the average value of the PPP interaction integrals in a given molecule. This is a useful limit, because as shown by Table 1, for all molecules except anthanthrene the conductance ratios calculated from $\tilde{G}$ correctly predict the direction in which the PPP ratio will deviate from the non-interacting magic ratio. Furthermore, the infinite-range interaction prediction is quantitatively correct within approx. 20%. Unfortunately, for anthanthrene with 3-12 and 9-22 connectivities, the infinite-range interaction limit conductance ratio is not a good approximation to the PPP ratio. We traced the latter failure to the fact that the Green's function element corresponding to the 9-22 connectivity crosses zero as a function of the interaction strength in the vicinity of the actual value of interaction. Therefore, the conductance ratio is very sensitive to the actual form and strength of interaction for this connectivity.

The orange dots in Fig. 4d are plots of HF conductance ratios versus those predicted by infinite range interaction model for all possible pairs of connectivities and show that there is a significant improvement compared with the non-interacting magic ratios. Moreover, as shown in Fig. 16 in the SI, typically the infinite-range interaction model correctly predicts in which direction the PPP conductance ratio will deviate from the non-interacting value.

The main result contained in Fig. 3, Fig. 4, and Table 1 is that the non-interacting conductance ratios are typically similar to those obtained in the presence of Coulomb interactions and therefore despite their simplicity, are a useful guide for predicting conductance ratios and identifying connectivities with high or low conductance. Furthermore for small molecules, where Lanczos results for the PPP model are available, the Lanczos ratios agree with those obtained using HF.

On the other hand, there are cases where interactions cause a strong deviation from non-interacting conductance ratios. We identified several pairs of connectivities for different molecules, where this is the case. For example, for anthanthrene we predict the conductance ratio for 6-7 and 1-10 connectivities to be about 275, which is much larger than the non-interacting ratio of 16 for this pair of connectivities. Additional examples are presented in Table 3 and 4 of the SI. These connectivities are interesting, because experimental measurement of their conductance ratios would establish that at least for certain connectivities, Coulomb interactions are needed to describe transport through such molecules.

**Discussion**

We have used exact (Lanczos) diagonalization, Hartree-Fock theory and density functional theory to examine conductance ratios of polyaromatic hydrocarbons with different connectivities to electrodes, which can be predicted using a simple and easy-to-use "magic number tables," such as those shown in Figs. 2c and 2d (and in Figs. 2-6 of the SI). We find that when Coulomb interactions and screening due to electrodes are switched on, conductance ratios are rather resilient, even though the conductances themselves vary. Consequently, although the precise numbers depend on the strength of the interaction and on screening, qualitative trends in conductance ratios can be predicted using non-interacting magic number tables. Overall the differences between HF, Lanczos and DFT predictions and variations due to screening are found to be comparable with deviations from experimental values. Therefore at the current level of experimental measurement, non-interacting magic numbers provide a useful tool for identifying molecules for subsequent experimental screening, without the need for large-scale computations involving electron-electron interactions. On the other hand, we have also identified examples where conductance ratios are sensitive to interactions. These molecules would be interesting targets for future synthesis, since their conductance ratios would demonstrate that in general both QI and interactions play an important role in controlling the flow of electricity through single molecules.

**Methods**

When analysing the PPP model, calculations for smaller molecules (benzene, naphthalene, and anthracene) are performed using both the Lanczos exact diagonalization method [35] and for larger molecules, where exact diagonalization is not feasible, we use the restricted Hartree-Fock (HF) approximation (for technical details see Methods and the SI). In both cases, the wide band approximation was used, in which the self energy due to the contacts is modelled by a single number. When including chemical details at an atomistic level, we use the SIESTA implementation of DFT combined with non-equilibrium Green's functions, in which the full self-energy matrix is computed.

This dual approach to modelling is needed, because correlated ab initio calculations with chemical specificity are not feasible. A similar combination of methods was utilised in [36], where in addition, the GW method was used. Within the PPP model, interactions are present within the molecule only, whereas interactions within the DFT mean-field treatment are present in both the molecules and electrodes.

**DFT-NEGF:** The optimized geometry and ground state Hamiltonian and overlap matrix elements of each structure was self-consistently obtained using the SIESTA implementation of density functional theory (DFT). SIESTA employs norm-conserving pseudo-potentials to account for the core electrons and linear combinations of atomic orbitals to construct the valence states. The generalized gradient approximation (GGA) of the exchange and correlation functional is used with the Perdew-Burke-Ernzerhof parameterization (PBE) a double-$\zeta$ polarized (DZP) basis set, a real-space grid defined with an equivalent energy cut-off of 250 Ry. [35, 36] The geometry optimization for each structure is performed to the forces smaller than 40 meV/Å. The mean-field Hamiltonian obtained from the converged DFT calculation or a simple tight-binding Hamiltonian was combined with Gollum quantum transport code [37] to calculate the phase-coherent, elastic scattering properties of the system consisting of left (source) and right (drain) leads and the scattering region. The transmission coefficient $T(E)$ for electrons of energy $E$ (passing from the source to the drain) is calculated via the relation $T(E) = Tr\{\Gamma_R(E)\mathcal{G}(E)\Gamma_L(E)\mathcal{G}^\dagger(E)\}$. In this expression, $\Gamma_{L,R}(E) = i\left(\Sigma_{L,R}(E) - \Sigma_{L,R}^\dagger(E)\right)$ describe the level broadening due to the coupling between left (L) and right (R) electrodes (which are modelled with atomic precision as shown in Fig. 5) and the central scattering region, $\Sigma_{L,R}(E)$ are the retarded self-energies associated with this coupling and $\mathcal{G} = (ES - H - \Sigma_L - \Sigma_R)^{-1}$ is the retarded Green's function, where $H$ is the Hamiltonian and $S$ is overlap matrix. Using obtained transmission coefficient $T(E)$, the conductance could be calculated by Landauer formula ($\sigma = \sigma_0 \int dE\, T(E)(-\partial f/\partial E)$) where $\sigma_0 = 2e^2/h$ is the conductance quantum, $f(E) = (1 + \exp((E - E_F)/k_B T))^{-1}$ is the Fermi-Dirac distribution function, $T$ is the temperature and $k_B$ is Boltzmann's constant.

**Hartree-Fock**: The PPP Hamiltonian, $H^{\text{int}} = \sum_{ijs} H_{ij}\, c_{is}^\dagger c_{js} + \frac{1}{2}\sum_{ij} U_{ij}(n_i - 1)(n_j - 1)$, contains matrix elements $H_{ij}$ of the non-interacting Hamiltonian (for the nearest-neighbour hopping integral we take $\gamma = 2.4\,\text{eV}$) and interaction integrals $U_{ij}$, which in the absence of screening by electrodes we calculate using the Ohno interpolation [33]: $U_{ij} = U/(1 + (U/(e^2/4\pi\varepsilon_0 d_{ij}))^2)^{-1/2}$ where $U = 11.13\,\text{eV}$ is the Hubbard parameter and $d_{ij}$ is the distance between sites $i$ and $j$. The interatomic distance is $d_0 = 1.4\,\text{Å}$. We take the image charge effects into account by analytically solving [38] the Poisson's equation for the electrostatic Greens function in a simplified geometry, namely we assume the electrodes are two infinite parallel plates located at a distance $d$ away from each of the connectivity sites. We decouple the interaction terms within the restricted HF approximation, yielding renormalized hopping matrix elements $H_{ij}^{\text{HF}} = H_{ij} - U_{ij}\langle c_{js}^\dagger c_{is}\rangle$. The expectation value is calculated from the Slater determinant built from the occupied scattering states of a molecule attached to electrodes. We model electrodes as tight-binding chains with nearest-neighbour hopping integral of $10\gamma$. The hopping integral between the connectivity site on the molecule and the nearest electrode site is $\gamma$, leading to coupling $\Gamma_{L,R} = 0.2\gamma$ (in the wide band limit we can neglect the energy dependence of $\Gamma$). The procedure is iterated until a self-consistent solution is obtained. Due to the chiral symmetry possessed by the PPP Hamiltonian of our molecules, the HF Hamiltonian has the same structure as the non-interacting one, i.e. the hopping integrals between atoms on the same sublattice as well as on-site energies remain zero. Once the convergence is achieved, the conductance is calculated with the Landauer-Büttiker formula [39,40] with the transmission function $T(E)$ read from the scattering state at energy $E$. We also performed unrestricted HF calculations where we allowed each sublattice to develop a magnetization. We found that the antiferromagnetic solution becomes the ground state only for interaction strengths that exceed the physically relevant ones by more than approx. 50%. For details, see the Supplementary information.

Systems with the chiral symmetry have a symmetric energy spectrum which means that for half-filled systems the Fermi energy is at the gap centre. [36] The chiral symmetry is defined and its consequences are explained in Supplementary Note 3. There it is shown that the PPP model as well as the corresponding Hartree-Fock Hamiltonian have this symmetry. Chiral symmetry ensures that the energy spectrum of the molecule is symmetric with respect to the centre of the HOMO-LUMO gap.

Clearly HF is an effective non-interacting theory, which creates new effective hoppings between non-neighbouring sites, which are absent from the non-interacting model. The inclusion of arbitrary long-range hoppings could significantly change the magic ratios, whereas those generated by the HF approximation using physically-relevant parameters do not.

**Lanczos**: The size of the Hilbert space grows exponentially with the size of the molecule and the exact full diagonalization of the PPP Hamiltonian is in our case limited to smallest system of benzene molecule. We therefore apply the Lanczos method [41], which allows for the treatment of larger systems as well as the calculation of ground state properties exactly. Within the Lanczos method one obtains the ground state $|\psi_0\rangle$ of an isolated molecule by starting from a random many-body state and then iteratively applying the Hamiltonian for generation of new basis states, within which the effective Hamiltonian is tridiagonal and easy to diagonalize. On the other hand, the core Green's function $G$ is obtained by starting the iterative procedure from $c_{is}|\psi_0\rangle$ or $c_{is}^\dagger|\psi_0\rangle$, and by calculating matrix elements between two series of Lanczos eigenstates for the Lehmann representation. The results converge within 80 iterative steps. The Green's function $\mathcal{G}$ of a molecule attached to electrodes is then calculated within the elastic co-tunneling approximation [42,43], i.e., the presence of the electrodes is taken into account with the Dyson's equation $\mathcal{G}^{-1} = G^{-1} - \Sigma_\text{L} - \Sigma_\text{R}$. The self-energies $\Sigma_\text{L,R}$ due to coupling to electrodes correspond to the same electrode-molecule couplings as in the HF calculation. The approximation is valid far from transmission resonances and above the Kondo temperature of the system. In our case both conditions are satisfied, because the Fermi level is at the centre of the HOMO-LUMO gap and there is no unpaired electron in the molecule. In the elastic co-tunneling approximation the conductance can again be calculated with the Landauer-Buttiker formula, with the transmission function obtained from $\mathcal{G}$ as $T(E) = \text{Tr}\{\Gamma_\text{R}(E)\mathcal{G}(E)\Gamma_\text{L}(E)\mathcal{G}^\dagger(E)\}$. [44]

**Data Availability**

All data generated or analysed during this study are included in this published article (and its Supplementary Information files).

**Acknowledgements**

L.U., T.R., J.K., and A.R. acknowledge the support of the Slovenian Research Agency under Contract No. P1- 0044. This work is supported by FET Open project 767187 – QuIET, the EU project BAC-TO-FUEL and the UK EPSRC grants EP/N017188/1, EP/N03337X/1 and EP/P027156/1.H.S. and S.S. acknowledge the Leverhulme Trust (Leverhulme Early Career Fellowships no. ECF-2017-186 and ECF-2018-375) for funding.

**Author Contributions**

C.J.L., A.R., T.R. and J.H.J. conceived and conducted the project. L.U. and T.R. carried out the Hartree-Fock formalism, J.K. contributed results with Lanczos technique and S.S. and H.S. carried out the DFT calculations. C.J.L., L.U. and T.R. wrote the manuscript. All authors took part in the discussions and reviewed the manuscript.

**Additional Information**

Supplementary information accompanies this paper
Competing Interests: The authors declare no competing interests.